\documentclass[12pt]{article}
\usepackage{psfig}
\usepackage{epsfig}
\usepackage{floatflt}
\newcommand\new{\newcommand}         
\newcommand\ren{\renewcommand}       
\ren{\textfraction}{0.01}            
\ren{\topfraction}{0.99}             
\ren\d{\mathrm{d}}                   
\catcode`\"=\active                  
\new"[1]{{\accent127#1}}             
\new\3{\ss }                         
\new{\mm}[1]{{\mbox{\hspace{#1mm}}}}
\new{\nin}{\noindent}
\new{\D}{\displaystyle}
\new{\mrm}{\mathrm}
\new{\eps}{\varepsilon}
\new\beq[1]{\begin{equation}{\label{eq:#1}}}
\new\eeq[0]{\end{equation}}
\new\barr[0]{\begin{array}}
\new\earr[0]{\end{array}}
\new\eq[1]{eq.(\ref{eq:#1})}
\new\Eq[1]{Equation(\ref{eq:#1})}
\new\fig[1]{fig.(\ref{fig:#1})}
\new\Fig[1]{Figure~(\ref{fig:#1})}
\new\tab[1]{tab.~(\ref{tab:#1})}
\new\Tab[1]{Table~(\ref{tab:#1})}
\new\app[1]{app.~(\ref{app:#1})}
\new\App[1]{Appendix~(\ref{app:#1})}
\new\fplace[3]{
\begin{minipage}[h]{#1\textwidth}
  \vspace*{#2}
  \begin{center}
      \mbox{\psfig{file=#3,width=\textwidth}}
  \end{center}
 \end{minipage}}
\new\cplace[4]{
\begin{minipage}[h]{#1\textwidth}
  \vspace*{#2}
  \caption[]{#4}
      \label{#3}
 \end{minipage}}
\new\av[1]{\left\langle #1 \right\rangle}
\new\ord[1]{\mbox{{\it O}$(#1)$}}
\new{\minus}{\mm{-4}-\mm{-4}}

\begin{document}
\begin{flushright}               
  MPI H - V17 - 2000 \\ 
  May 31, 2000
\end{flushright}

\vspace*{3mm}
\begin{center}  
 {\bf\Large Averaging Measurements with Hidden Correlations}\\[2mm] 
 {\bf\Large and Asymmetric Errors}\\[5mm]
 {\large Michael Schmelling / MPI for Nuclear Physics} \\[2mm]
 {\large Postfach 103980, D-69029 Heidelberg}\\[5mm]
\end{center}

\begin{quote}
\begin{center}
\subsection*{Abstract}
\end{center}
  Properties of weighted averages are studied for the general 
  case that the individual measurements are subject to hidden 
  correlations and have asymmetric statistical as well as
  systematic errors. Explicit expressions are derived for an
  unbiased average with a well defined error estimate. 
\end{quote}

\vspace*{1mm}
\section{Introduction}
Combining independent measurements of a physical quantity in order to summarize 
the information from different sources into one number is usually done by 
calculating a weighted average, where each measurement contributes with a 
weight which is inversely proportional to its variance. 

In practical applications, however, one is often confronted with the problem 
that the measurement errors which are quoted define a 68\% confidence level 
interval around a Maximum Likelihood estimate, whereas one would need an 
unbiased estimator with its standard deviation in order to form a weighted 
average. In addition, one usually has to deal with statistical and systematic 
uncertainties, and different results may be correlated, e.g. due to common 
uncertainties in some theoretical prediction, while a quantitative estimate 
for the size of the correlation does not exist. 

This paper presents a systematic approach to form a meaningful weighted average 
under such conditions. After a short reminder of the properties of weighted 
averages, it first is shown how to deal with hidden correlations in the data.
Then a detailed analysis of parameter estimates based on the Maximum Likelihood 
Method is performed, and the consequences when forming a weighted average are 
studied in detail. The results are illustrated by means of a numerical simulation.
Finally, the issue of systematic errors and how to combine statistical and systematic
errors is addressed.

\section{Weighted Averages}
Let $\{x_i\}$\/ be a set of $n$\/ unbiased measurements of a parameter $\mu$, with 
expectation values $\av{x_i}$\/ and covariance matrix $C_{ij}(x)$\/ given by
\beq{nobiascx}
    \av{x_i} = \mu
    \mm{5} \mbox{and} \mm{5}
     C_{ij}(x) = \av{(x_i-\mu)(x_j-\mu)} = \av{x_i x_j}  - \mu^2.
\eeq
From $\{x_i\}$\/ a new estimator $\bar{x}$\/ for $\mu$\/ can be constructed, 
which has a smaller variance than any of the $x_i$, by minimizing
\beq{chisq}
   \chi^2 = \sum_{i,j} (x_i-\bar{x}) (x_j-\bar{x}) C^{-1}_{ij}(x).
\eeq
Note that the $\chi^2$-function is constructed in such a way, that it is invariant 
with respect to arbitrary linear transformations of the variables with a regular 
transformation matrix. In case the individual $x_i$\/ are independent, the covariance 
matrix becomes diagonal, $C_{ij}(x)= \sigma_i^2 \delta_{ij}$\/ and \eq{chisq} 
simplifies to
\beq{chisqdiag}
   \chi^2 = \sum_i \frac{(x_i-\bar{x})^2}{\sigma_i^2} .
\eeq
The value $\bar{x}$\/ which minimizes $\chi^2$\/ and thus in the least squares 
sense is optimally consistent with all measurements $x_i$\/ is found to be
\beq{minchisq}
   \bar{x} = \sum_i w_i x_i 
   \mm{5} \mbox{with} \mm{5}
   w_i = \frac{1/\sigma^2_i}{N}
   \mm{5} \mbox{and} \mm{5}
   N = \sum_k \frac{1}{\sigma^2_k},
\eeq
and the variance of $\bar{x}$\/ is obtained by standard error propagation as 
\beq{varmubar}
   C(\bar{x})  = \sigma^2(\bar{x})
               = \sum_{ij} w_i w_j C_{ij}
               = \sum_i w_i^2 \sigma^2_i 
               = N^{-1}.
\eeq
Another relevant quantity is the expectation value of $\chi^2$\/ at the 
minimum
\beq{chi2av}
  \av{\chi^2_{\min}} = \sum_i \av{\frac{(x_i-\bar{x})^2}{\sigma_i^2}} 
              = \sum_i \frac{\av{x^2_i}}{\sigma^2_i}  - N \av{\bar{x}^2}
              = n - 1
\eeq
Note that although the value of $\av{\chi^2}$\/ has been derived for the case 
of a diagonal covariance matrix, the result is valid in general since one always 
can find a linear transformation of variables which diagonalizes $C_{ij}$\/  
while $\chi^2$\/ by construction is invariant.

Minimizing $\sum_i w_i^2 \sigma_i^2$\/ with respect to $w_i$\/ subject to the 
constraint $\sum_i w_i=1$, one sees that the weights $w_i$\/ used for the 
weighted average \eq{minchisq} lead to the smallest possible variance for 
$\bar{x}$\/ if the individual measurements are uncorrelated. 

In case that correlations are present, one can still calculate a weighted average 
according to \eq{minchisq}, which always is an unbiased estimator for $\mu$, even
if it no longer is the value which minimizes \eq{chisq}. It also does not have 
the smallest possible variance, although the loss in precision usually is only
marginal. In practical applications one thus might prefer the robust average
\eq{minchisq} to the formally optimal result from minimization of \eq{chisq}. As 
this requires to calculate the inverse of $C_{ij}$, it can only be performed
reliably if the off-diagonal elements of $C_{ij}$\/ are known precisely. Otherwise 
numerical instabilities are likely to render the formally optimal result $\bar{x}$\/ 
meaningless. Note, however, that while it is perfectly meaningful to calculate 
a weighted average ignoring off-diagonal elements of the covariance matrix, the
full matrix $C_{ij}$\/ must be taken into account for the variance of the average.

For uncorrelated inputs the weighted average \eq{minchisq} is 
the optimal way to combine measurements in order to minimize the variance of 
the result. It is independent of the actual shape of the probability density
functions (PDFs) involved. All one needs to know are the values $x_i$\/ and the 
variances $\sigma^2_i$\/ of the individual measurements. Note that the 
confidence levels associated to the variances may vary with the shape of the 
PDF. The weighted average, on the other hand, has a PDF which is a convolution 
of the PDFs of the individual $x_i$, and according to the central limit 
theorem with an increasing number of contributions quickly converges towards 
a Gaussian, where the range $\bar{x}\pm\sigma(\bar{x})$\/ corresponds 
approximately to an 68\% confidence level interval.

\section{Hidden Correlations}
In practical applications one sometimes is confronted with a situation where
measurements have to be averaged which are suspected to be correlated, but where
the correlations are very hard to quantify. While, as argued before, a weighted
average can still be calculated, it is not obvious how to give a meaningful 
estimate for the error of the estimate. One possible approach to this problem
has been discussed in \cite{ref:msav}. The basic idea was to use the value of the 
$\chi^2$\/ calculated according to \eq{chisq} as an indicator and adjust a global 
correlation coefficient between all measurements such that it becomes equal to
its expectation value. Here a similar argument will be presented, which is 
more convenient as it leads to a much simpler scaling prescription.

A value $\chi^2$\/ larger than its expectation value can result if anti-correlations 
are present, or if the measurement errors are underestimated. In case of the latter, 
$N^{-1}$\/ from \eq{varmubar} underestimates the variance of $\bar{x}$, in the former 
case it is overestimated. In order to be conservative, a common practice~\cite{ref:pdg} 
is to assume that a large $\chi^2$\/ is a consequence of too small measurement errors 
and apply a common scaling factor $\chi^2/(n-1)$\/ to all variances to obtain
a $\chi^2$\/ identical to its expectation value. Although one certainly can 
question this approach, it has to be pointed out, that unless there are severe 
discrepancies in the data, this procedure still gives an error estimate for $\bar{x}$\/ 
which is smaller than any of the individual errors $\sigma_i$\/ of the data. 

The other indication for a possible problem is a too small value for $\chi^2$.
This can either mean overestimated measurement errors or positive correlations
between the data. While the former would imply that the error of $\bar{x}$\/ 
in reality is smaller than $N^{-1}$, the latter would require a larger error. 
Being conservative, again an estimate for the larger error shall be derived. As 
these considerations only affect error estimates, the calculations can be simplified
without loss of generality by assuming $\av{x_i}=\av{\bar{x}}=0$. This implies
$\av{x^2_i}=\sigma^2_i$\/ and $\av{\bar{x}^2}=C(\bar{x})$, and the expectation 
value $\av{\chi^2_{\min}}$\/ \eq{chi2av} simplifies to
\beq{avc}
    \av{\chi^2_{\min}} = n - N C(\bar{x}).
\eeq
If the individual measurements are uncorrelated, then one has $C(\bar{x})=N^{-1}$,
and as shown before $\av{\chi^2_{\min}} = n-1$. If there are reasons to suspect that the 
measurements $x_i$\/ are not independent, then one can re-interpret \eq{avc}, assuming 
that the actual best fit $\chi^2$-value is equal to its expectation value and derive a 
variance estimate 
\beq{cscale}
   C(\bar{x}) = \frac{1}{N} (n - \chi^2) .
\eeq
This expression constitutes a simple scaling prescription for the case that the 
$\chi^2$\/ of a weighted average is smaller than expected. Like the scaling procedure 
for the errors in case of too large $\chi^2$\/ values, it defines an effective way
to take imperfections or incomplete knowledge about the input data into account.

Like the weighted average, also the simple scaling~\eq{cscale} of the uncorrelated 
variance estimate $N^{-1}$\/ is readily generalized to $m$-dimensional vector-valued 
measurements $\vec{x}_i$. It will also be demonstrated explicitly that \eq{cscale} 
is equivalent to the assumption of a fixed common covariance between all 
measurements. In contrast to this, the more complicated procedure discussed in 
\cite{ref:msav} was based on the assumption of a common correlation coefficient. In
practice both scaling schemes give rather similar results. Numerically the method
\cite{ref:msav} yields slightly smaller errors, indicating that the assumption of 
a common correlation coefficient tends to assign a higher weight for more precise
measurements than the assumption of a common covariance.

Ignoring correlations between measurements, the generalization of \eq{chisqdiag} 
to vector valued measurements $\vec{x}_i$\/ with covariance matrices $C_i$\/ 
is given by
\beq{chisqvmin}
    \chi^2 
  = \sum_i \left(\vec{x}_i-\bar{x}\right)^T C^{-1}_i \left(\vec{x}_i - \bar{x}\right).
\eeq
Note that in this expression the index $i$\/ enumerates entire vectors rather than
individual components. Those vectors $\vec{x}_i$\/ are assumed to be uncorrelated; 
correlations between the components of $\vec{x}_i$\/ are described by the covariance
matrices $C_i$. The minimum of \eq{chisqvmin} is obtained for 
\beq{subst}
  \bar{x} = N^{-1} \sum_i C^{-1}_i \vec{x}_i 
  \mm{5} \mbox{with} \mm{5}
  N = \sum_i C^{-1}_i.
\eeq
Note that now $\bar{x}$\/ is a vector of dimension $m$. Generalizing \eq{chi2av} the 
value of $\chi^2$\/ at the minimum then can be expressed as
\beq{chi2minv}
   \chi^2_{\min} 
 = \mbox{Tr} \left( \sum_i C_i^{-1} \vec{x}_i\vec{x}_i^T - N \bar{x}\bar{x}^T \right)
\eeq
Without loss of generality one can again make the simplifying assumption 
$\av{\vec{x}_i}=\av{\bar{x}}=0$. This leads to $\av{\vec{x}_i\vec{x}_i^T}=C_i$\/ and
$\av{\bar{x}\bar{x}^T}=C(\bar{x})$, and the expectation value of $\chi^2_{\min}$\/ 
becomes
\beq{chi2minex}
   \av{\chi^2_{\min}} 
 = n \cdot m - \mbox{Tr} \left(N \av{\bar{x}\bar{x}^T}\right)
 = n \cdot m - \mbox{Tr} \left(N C(\bar{x}) \right). 
\eeq
At this point one could directly apply the same short argument as before and derive 
the generalization of \eq{cscale} for $m$-dimensional measurements. However, in 
order to illuminate a little bit the background of the scaling rule, here a different 
route shall be taken, making the explicit assumption that the correlation between 
the vectors $\vec{x}_i$\/ is such, that the covariance matrix $C_i$\/ for each 
vector $\vec{x}_i$\/ is the sum of a common contribution $C$\/ and a specific 
component $S_i$
\beq{defcorr}
    C_i = C + S_i 
    \mm{5} \mbox{i.e.} \mm{5}
    \av{\vec{x}_i \vec{x}^T_j} 
    = \left\{ \begin{array}{ll}  C    & \mbox{for}\mm{2} i\neq j \\
                                 C_i  & \mbox{for}\mm{2} i = j 
              \end{array}
      \right.
\eeq 
One then obtains
\beq{covy}
  C(\bar{x}) = \av{\bar{x}\bar{x}^T} 
   = N^{-1} \left[ {\bf 1} + N C - \left( \sum_i C^{-1}_i C C^{-1}_i \right) N^{-1} \right],
\eeq
where $\bf{1}$\/ denotes the unit matrix, and 
\beq{chi2minex2}
   \av{\chi^2_{\min}} = m\cdot(n-1)
 - \mbox{Tr} \left[ N C - \left( \sum_i C^{-1}_i C C^{-1}_i \right) N^{-1}
             \right].
\eeq
For $C=0$\/ the covariance matrix of the average is given by $N^{-1}$\/ and one has
$\av{\chi^2_{\min}} = m\cdot(n-1)$. A significant deviation of the $\chi^2_{\min}$\/
from this expectation value can be taken as an indication for correlations $C$\/ 
between the individual measurements. The simplest way to take these correlations into 
account for the error of the average, is by scaling the uncorrelated estimate $N^{-1}$\/ 
by a factor $(1+c)$. From \eq{covy} then follows that $C$\/ satisfies the relation 
\beq{defcond}
   N C - \left( \sum_i C^{-1}_i C C^{-1}_i \right) N^{-1} = c\; {\bf 1}.
\eeq
For a given $c$\/ this is a linear equation which can be solved numerically for $C$\/
by iterating for example the fixed point condition
\beq{fixedpoint}
   C_{n+1} = N^{-1} + N^{-1} \left( \sum_i C^{-1}_i C_n C^{-1}_i \right) N^{-1}
\eeq
and setting $C= c\;C_{\infty}$. In practical applications this only has to be done
if one wants to extract $C$\/ in order to get a quantitative estimate for the size
of the correlations between the measurements. Otherwise it is sufficient to fix the 
value of the parameter $c$\/ by requiring the observed $\chi^2$-value to be 
equal to the expectation value $\av{\chi^2_{\min}}$. With \eq{defcond} and 
\eq{chi2minex2} one obtains
\beq{def}
  \chi^2_{\min} = m(n-1) - m\;c 
\eeq
and a scaled variance estimate
\beq{scalvar}
   C(\bar{x}) = N^{-1} \left( n - \frac{\chi^2}{m} \right).
\eeq
For $m=1$\/ the result \eq{cscale} is recovered.

To summarize, unless correlations between measurements are really well understood, 
it appears advisable to employ the weighted average ignoring all correlations as a 
robust procedure to combine several inputs into one number. Care, however, has to 
be exercised to give a realistic error estimate. In the absence of other information 
the $\chi^2$\/ of the average is a useful indicator for potential problems. The 
conservative approach would be to interpret a $\chi^2$\/ smaller than its expectation 
value as evidence for positive correlations between the measurements, and a $\chi^2$\/ 
larger than its expectation value as evidence that the errors are underestimated. In 
both cases the simple estimate $N^{-1}$\/ for the variance of the weighted average would 
have be to increased, which is conveniently done according to the following scheme
\beq{recipe}
    C(\bar{x}) 
  = N^{-1} \left\{ \begin{array}{c} 
                     \displaystyle
                     \left( n - \frac{\chi^2}{m} \right) 
                     \mm{5} \mbox{for} \mm{5} \chi^2 \leq m(n-1) \\
                     \displaystyle
                     \frac{\chi^2}{m(n-1)}
                     \mm{7} \mbox{for} \mm{5} \chi^2 > m(n-1)
                   \end{array}
           \right.         
\eeq
It has to be emphasized, that such a scaling should only be performed if 
one has reasons to believe that the variance estimate $N^{-1}$\/ of the weighted 
average underestimates the true uncertainties, or if one wants to quote a 
conservative error. Otherwise, since the $\chi^2$-distribution especially for 
a small number of degrees of freedom has a large relative width, one will 
systematically bias the error to large values.

\section{Statistical Errors}
In order to form a weighted average of independent data one needs unbiased 
measurements with known variances. On the other hand, in order to quantify
the statistical precision of a measurement, usually a 68\% confidence
level interval is quoted, which is constructed such that the true value of 
the parameter is inside this interval in 68\% of all cases. For a Gaussian
distribution the 68\% confidence level interval is given by $\bar{x}\pm\sigma$,
where $\bar{x}$\/ is the mean value and $\sigma$\/ the rms-width of the 
distribution, i.e. the square of the error is the variance. In general one
needs to know the shape of the PDF in order to extract the variance from the 
error interval. 

If the shape of the PDF is not explictly specified, one has to resort to 
certain reasonable assumptions in order to proceed. For the following it 
will be assumed that in absence of other information the primary result of 
a measurement has a Gaussian PDF. Using the Maximum Likelihood Method this
measurement then is interpreted in terms of a physical parameter, with a 
nominal result and an error range covering a 68\% confidence level interval.  
Under these conditions a rigorous interpretation can be attached to asymmetric 
errors even if no further information about the PDF is given.

\subsection{Interpretation of Asymmetric Errors}
As explained above, a measurement $x$\/ is assumed to scatter around a 
mean value $\av{x}$\/ according to a Gaussian PDF with variance $\sigma^2$.
The precision of the measurement $\sigma$\/ is assumed to be known, e.g. from 
first principles or some calibration procedure. The mean value $\av{x}$\/ is
related to a physical parameter $\mu$\/ via a function $s(\mu)$, $\av{x} = s(\mu)$. 
The likelihood function $p(x|\mu)$\/ to observe a value $x$\/ given $\mu$, then is
\beq{pxmu}
   p(x|\mu) = \frac{1}{\sqrt{2\pi}\sigma}
              \exp\left( -\frac{(x-s(\mu))^2}{2\sigma^2} \right).
\eeq         
In the framework of the Maximum Likelihood Method the function \eq{pxmu} is used 
to extract an estimate $\mu_{ml}$\/ for $\mu$\/ from a given measurement $x$\/ as 
the parameter which maximizes the likelihood. The solution is evidently given by 
$\mu_{ml}=s^{-1}(x)$. Note that the likelihood function here is merely a tool to 
extract an estimate $\mu_{ml}$\/ for $\mu$\/ by looking for the value which 
maximizes $p(x|\mu)$, i.e. no attempt is made to interpret the likelihood function 
as a PDF for $\mu$. Note that in a Bayesian approach using a flat prior 
distribution, one could actually consider \eq{pxmu} as a probability density for 
$\mu$\/ and take e.g. the average $\av{\mu}$\/ of $p(x|\mu)$\/ instead of $\mu_{ml}$\/
as an estimate for $\mu$, although, as shown below, the Maximum Likelihood estimate 
$\mu_{ml}$\/ usually is the better choice.

In addition to the parameter estimate also an error interval has to be given. 
This is done by translating the error interval $[x_-,x_+]$\/ of the measurement 
to an error interval for the Maximum Likelihood estimate according to 
$[\mu_-,\mu_+]=[s^{-1}(x_-),s^{-1}(x_+)]$. By conservation of probability both 
intervals have the same probability content. If $p(x)$\/ denotes the PDF for the 
measurement $x$, then the PDF for the Maximum Likelihood estimate $q(\mu_{ml})$\/ 
is given by
\beq{pmuml}
     q(\mu_{ml}) = \int_{-\infty}^{\infty} dx\; p(x)\; \delta(\mu_{ml}-s^{-1}(x))
\eeq
and one obtains for the probability content
\beq{probmuhat}
    \int_{\mu_-}^{\mu_+} d\mu_{ml}\; q(\mu_{ml})
  = \int_{s^{-1}(x_-)}^{s^{-1}(x_+)} d\mu_{ml}\; 
    \int_{-\infty}^{\infty} dx\; p(x)\; \delta(\mu_{ml}-s^{-1}(x))
  = \int_{x_-}^{x_+} dx\; p(x).
\eeq
Usually the error interval for the measurement is given by 
$[x_-,x_+]=[x-\sigma,x+\sigma]$\/ which in case of a Gaussian PDF includes 
the expectation value $\av{x}$\/ in approximately 68\% of all cases. The 
same confidence level then also applies for $[\mu_-,\mu_+]$.
From \eq{pxmu} one sees that this definition of error interval corresponds to 
finding regions where the logarithm of the likelihood function stays within half
a unit of the maximum, i.e. it is the shortest interval which can be constructed
for a given probability content. This concludes the discussion of what will be 
assumed to be the connection between a measurement $x\pm\sigma$\/ and the 
corresponding Maximum Likelihood parameter estimate $\mu_{ml}\pm\sigma_{\pm}$.

It is now possible to study the expectation value and variance of $\mu_{ml}$. 
With \eq{pmuml} the expectation value becomes 
\beq{avmuhat}
    \av{\mu_{ml}} 
  = \int d\mu_{ml} \; \mu_{ml} \; q(\mu_{ml})
  = \int dx \; p(x) \; s^{-1}(x).
\eeq
It is clearly not possible to determine this expectation value for arbitrary functions
$s^{-1}(x)$, but one can address the problem in a systematic way by expanding 
$s^{-1}(x)$\/ around $\av{x}=s(\mu)$. Since an asymmetric error will not come about for 
linear functions $s(\mu)$, the expansion should at least go to second order. Assuming 
that for practical applications the non-linearities will be small, the expansion will 
be truncated there. 
\beq{exps}
    s^{-1}(x) = \mu + \alpha y + \beta y^2
    \mm{5} \mbox{with} \mm{5}
    y = x - \av{x} .
\eeq
Substituting this into \eq{avmuhat} and using $\av{y^2}=\sigma^2$\/ then yields
\beq{mubias}
    \av{\mu_{ml}} = \mu + \beta\sigma^2.
\eeq
One finds that the maximum likelihood estimate $\mu_{ml}$\/ is a biased estimator,
with a constant bias proportional to the curvature of $s^{-1}(x)$. An unbiased 
estimator can be constructed by exploiting the information contained in the asymmetric 
errors. As discussed before, the error range $\mu_{\pm}=\mu_{ml}\pm\sigma_{\pm}$\/ 
is defined by the condition $\mu_{\pm}=s^{-1}(x\pm\sigma)$, which yields
\beq{sigmapm}
   \sigma_{+} = \alpha\sigma + 2\beta\sigma y + \beta\sigma^2
   \mm{5} \mbox{and} \mm{5}
   \sigma_{-} = \alpha\sigma + 2\beta\sigma y - \beta\sigma^2.
\eeq
The difference of the two is proportional to the bias term in \eq{mubias} so 
that an unbiased estimator $\hat{\mu}$\/ for $\mu$\/ is given by
\beq{mutrue}
   \hat{\mu} = \mu_{ml} - \frac{1}{2}(\sigma_+ - \sigma_-) .
\eeq
At first glance this result may be surprising, since it implies that in case the 
positive error is larger than the negative one, the unbiased estimate $\hat{\mu}$\/ 
for the parameter is in fact smaller than $\mu_{ml}$. This underlines the fact that 
the asymmetric error does not describe a likelihood function for $\mu$. Instead, 
a large positive error means that the value $\mu_{ml}$\/ on average will 
overestimate the true value, which is then compensated by subtracting half the 
difference of the errors.
 
In terms of the expansion \eq{exps} the unbiased estimate $\hat{\mu}$\/ is given by 
\beq{munbiased}
  \hat{\mu} = \mu + \alpha y + \beta y^2 - \beta\sigma^2
\eeq
and with
\beq{momgauss}
   \av{y^n} 
 = \frac{1}{\sqrt{2\pi}\sigma}
   \int dy y^n \exp\left(-\frac{y^2}{2\sigma^2}\right)
 = \left\{ 
   \begin{array}{ll}
     0 & n = 2k-1 \\
     1\cdot 3 \cdots (2k-1) \sigma^{2k} & n = 2k
   \end{array}
   \right. 
\eeq
one finds for the variance of $\hat{\mu}$ 
\beq{varmuhat}
   C(\hat{\mu}) = \alpha^2\sigma^2 + 2\beta^2\sigma^4 .
\eeq

From $\hat{\mu}$\/ and the asymmetric errors alone it is not possible to determine 
the value of the variance, since in addition to $\mu$, $\alpha$\/ and $\beta$\/ also 
$(x-\av{x})$\/ is an unknown, i.e. one is faced with three constraints for four 
variables. In the same spirit, however, as $\hat{\mu}$\/ is an unbiased estimator for 
$\mu$, one can also construct an unbiased estimator $\hat{C}(\hat{\mu})$\/ for $C$. 
The solution is given by 
\beq{varhat}
   \hat{C}(\hat{\mu}) 
 = \frac{1}{4}(\sigma_+ + \sigma_-)^2 - \frac{1}{2}(\sigma_+ - \sigma_-)^2 .
\eeq
One easily verifies that the expectation value of \eq{varhat} reproduces \eq{varmuhat}.
The functional expression \eq{varhat} can be interpreted as the naive estimate for 
the variance of $\hat{\mu}$\/ which is reduced by a correction term that vanishes
for symmetric errors. An important aspect is that $\hat{C}(\hat{\mu})$\/ is only 
an estimate for the variance of $\hat{\mu}$, even if the variance $\sigma^2$\/ of
the measurement is known precisely. Only for $\beta=0$, i.e. when the connection 
between measurement $x$\/ and parameter $\mu$\/ is linear one recovers $\hat{C}=C$.
As shown below, the fact that $\hat{C}(\hat{\mu})$\/ is subject to statistical 
fluctuation complicates matters a lot.

\subsection{Weighted Averages with Asymmetric Errors}
\label{sec:wavasym}
In the previous subsection a prescription was derived to quote an unbiased 
estimate for mean value and variance of a parameter $\mu$. In terms of the 
expansion \eq{exps} the results are
\beq{collect}
 \hat{\mu} = \mu + \alpha y + \beta y^2 - \beta\sigma^2 
 \mm{5} \mbox{and} \mm{5}
 \hat{C}  = (\alpha\sigma + 2\beta\sigma y)^2 - 2\beta^2\sigma^4 .
\eeq
With \eq{varmuhat} this can be rewritten as
\beq{varbias}
  \hat{C} - C = \gamma (\hat{\mu} - \mu)
  \mm{5} \mbox{with} \mm{5}
  \gamma =  4\beta\sigma^2 = 2(\sigma_+ - \sigma_-).
\eeq
It follows immediately that the variance of $\hat{C}$\/ is given by $\gamma^2 C$,
i.e. it vanishes for symmetric errors. In case of asymmetric errors both $\hat{C}$\/ 
and $\hat{\mu}$\/ are random variables which scatter around their respective 
expectation values, and according to \eq{varbias} both are fully correlated.
Although individually both variables are unbiased, their ratio $\hat{\mu}/\hat{C}$\/
is not. As a consequence also the weighted average $\bar{\mu}$\/ 
\beq{wavbar}
   \bar{\mu} 
 = \frac{\sum_i \hat{\mu}_i/\hat{C}_i} {\sum_i 1/\hat{C}_i}
 = \mu +  \frac{\sum_i (\hat{\mu}_i-\mu)/\hat{C}_i} {\sum_i 1/\hat{C}_i}.
\eeq
is no longer an unbiased estimator for the true parameter value $\mu$. Introducing 
\beq{defdc}
   d_i = \hat{\mu}_i - \mu 
   \mm{5} \mbox{and} \mm{5}
   \hat{C}_i = C_i + \gamma_i d_i ,
\eeq
the expectation value of $\bar{\mu}$\/ can be written as
\beq{avwavbar}
  \av{\bar{\mu}} = \mu + 
  \left\langle \frac{\sum_i d_i/(C_i + \gamma_i d_i)}{\sum_i 1/(C_i+\gamma_i d_i)}
  \right\rangle
  = \mu + \left\langle \frac{Z_1}{Z_0} \right\rangle
  \mm{5} \mbox{with} \mm{5}
  Z_m = \sum_i \frac{d_i^m}{C_i+\gamma_i d_i} ,
\eeq
and the bias of the weighted average is given by the expectation value of the 
ratio $Z_1/Z_0$. 

Using a Taylor expansion about the origin, the bias $\av{Z_1/Z_0}$\/ can be 
expressed through the moments of the deviates $d_i$. For a general function $F$\/ 
of deviates $d_i$\/ one has
\beq{expf}
  \left\langle F \right\rangle
 = \sum_{n=0}^{\infty} \frac{1}{n!} 
  \left\langle \left( \sum_k d_k \frac{\partial}{\partial d_k} \right)^n F(0) 
  \right\rangle .
\eeq
The notation $F(0)$\/ implies that all derivatives of $F$\/ are to be taken at 
$d_i=0$. For the problem at hand the first moments are zero, $\av{d_i}=0$, and 
different deviates are independent, $\av{d_k^m d_l^n} = \av{d_k^m}\av{d_l^n}$\/ for
$k \neq l$. Under these conditions \eq{expf} simplifies considerably and the 
leading order terms become
\beq{expfterms}
    \left\langle F \right\rangle 
  = \left( 1 + \sum_k \frac{\av{d_k^2}}{2!} \frac{\partial^2}{\partial d_k^2}
             + \sum_k \frac{\av{d_k^3}}{3!} \frac{\partial^3}{\partial d_k^3}
             + \sum_k \frac{\av{d_k^4}}{4!} \frac{\partial^4}{\partial d_k^4}
             + \sum_{k<l} \frac{\av{d_k^2}}{2!} \frac{\av{d_l^2}}{2!} 
               \frac{\partial^2}{\partial d_k^2} \frac{\partial^2}{\partial d_l^2}
             \ldots
    \right) F(0)  . 
\eeq
The evaluation of the moments $\av{d_i^n}$\/ using \eq{momgauss} is straightforward.
The leading order results, expressed as function of the variance $C_k$\/ and 
the asymmetry parameter $\gamma_k$, are
\beq{dkmoments}
  \av{d_k^2} = C_k 
  ,\mm{5}
  \av{d_k^3} = \frac{3}{2} C_k\gamma_k - \frac{1}{16}\gamma_k^3 
  \mm{5}\mbox{and}\mm{5}
  \av{d_k^4} = 3 C_k^2 + 3\gamma_k^2 C_k - \frac{3}{16}\gamma_k^4   .
\eeq
At this points all ingredients are available which are needed to construct an 
unbiased weighted average also in presence of asymmetric errors. To simplify the 
organization of terms it is convenient to define, in addition to $Z_m$\/ 
from~\eq{avwavbar}, the following sums:
\beq{deftlm}
   T_{lm} = \sum_k \frac{\gamma_k^l}{(C_k + \gamma_k d_k)^m}
   \mm{5} \mbox{and} \mm{5}
   S_{lm} = \sum_k \frac{\gamma_k^l}{C_k^m} = T_{lm}(0) .
\eeq
For the derivatives of $Z_m$\/ and $T_{lm}$, $n>0$, one finds
\beq{dtlm}
    \frac{\partial^n}{\partial d_k^n} Z_m(0)
  = n! \frac{(-\gamma_k)^{n-m}}{C_K^{n-m+1}}    
  \mm{2} \mbox{if} \mm{2} n\geq m \mm{5} \mbox{and} \mm{5}
    \frac{\partial^n}{\partial d_k^n} T_{lm}(0)
  = \frac{(-\gamma_k)^{n+l}}{C_k^{n+m}} \frac{(n+m-1)!}{(m-1)!}.
\eeq
Evidently every derivative picks up one power of $\gamma_k$\/ and the expansion 
around $d_i=0$\/ will be a power series in $\gamma_i$. The error of a truncated
series will be quoted as \ord{\gamma^n}, meaning a sum where each term is a product 
of at least $n$\/ factors $\gamma_i$. In general these factors will come from 
different measurements $\hat{\mu}_i$.

The actual calculations are straightforward but lengthy. Results of auxiliary 
calculations are collected in the appendix. For the bias on the weighted average 
one finds
\beq{sumbias}
     \left\langle \frac{Z_1}{Z_0} \right\rangle
  = -\frac{S_{11}}{S_{01}} + \frac{S_{12}}{S_{01}^2} + \ord{\gamma^3} .
\eeq
\Eq{sumbias} shows that in case of asymmetric
error the bias is of \ord{\gamma}, i.e. proportional to the asymmetry of the 
errors. After correction for the dominant effect the residual bias is only of 
\ord{\gamma^3}, because both the third derivative of $Z_1/Z_0$\/ and the 
third moments $\av{d_k^3}$\/ each provide one extra power of $\gamma$. The 
higher order terms also contribute at least two factors $\gamma$.

In practical applications a bias correction according to \eq{sumbias} will be 
done by substituting the $T_{lm}$\/ for the $S_{lm}$. This is permitted as long 
as the expectation value $\av{T_{lm}}$\/ is equal to $S_{lm}$\/ to the same 
order in $\gamma$\/ as the terms that are anyhow neglected in the correction 
terms. As shown in the appendix, the additional bias from this substitution for 
\eq{sumbias} is again of \ord{\gamma^3}. To leading order it thus is consistent 
to use $T_{lm}$\/ instead of $S_{lm}$\/ to obtain a bias corrected weighted 
average $\bar{\mu}_c$ 
\beq{wavcorr}
   \bar{\mu}_c 
 = \bar{\mu} + \frac{T_{11}}{T_{01}} - \frac{T_{12}}{T^2_{01}}
 = \mu + \frac{Z_1}{Z_0} + \frac{T_{11}}{T_{01}} - \frac{T_{12}}{T^2_{01}}
 = \mu + \Delta
\eeq
with $\bar{\mu}_c-\mu=\ord{\gamma^3}$. 

The variance $C(\bar{\mu}_c)$\/ of the bias corrected average $\bar{\mu}_c$\/ 
is given by 
\beq{varmuc}
\begin{array}{lcccccccr} 
     C(\bar{\mu}_c) =  \av{\Delta^2} - \av{\Delta}^2 
 &   =    & \av{\frac{Z^2_1}{Z^2_0}} 
 & \minus & \av{\frac{Z_1}{Z_0}}^2  
 &  +     &  2 \left( \av{\frac{Z_1}{Z_0} \frac{T_{11}}{T_{01}}} \right. 
 & \minus & \left. \av{\frac{Z_1}{Z_0}} \av{\frac{T_{11}}{T_{01}}} \right) \\
 &  +     & \av{\frac{T^2_{11}}{T^2_{01}}}
 & \minus & \av{\frac{T_{11}}{T_{01}}}^2
 &  -     & 2 \left( \av{\frac{Z_1}{Z_0} \frac{T_{12}}{T^2_{01}}} \right. 
 & \minus & \left. \av{\frac{Z_1}{Z_0}} \av{\frac{T_{12}}{T^2_{01}}} \right) \\
 &  +     & \av{\frac{T^2_{12}}{T^4_{01}}}
 & \minus & \av{\frac{T_{12}}{T^2_{01}}}^2 
 &  -     & 2 \left( \av{\frac{T_{11}}{T_{01}} \frac{T_{12}}{T^2_{01}}} \right. 
 & \minus & \left. \av{\frac{T_{11}}{T_{01}}} \av{\frac{T_{12}}{T^2_{01}}} \right) .
\end{array}
\eeq
The individual contributions to this expression can be found in the appendix. 
Collecting all terms one obtains
\beq{varmucexp}
    C(\bar{\mu}_c) 
 = \frac{1}{S_{01}}  + \frac{S_{22}}{S^2_{01}}
 - \frac{2 S_{23}}{S^3_{01}} + \frac{S^2_{12}}{S^4_{01}}
 + \ord{\gamma^4} .
\eeq
In practical applications one again has to use $T_{lm}$\/ instead of $S_{lm}$,
which for the first term in \eq{varmucexp} introduces an additional bias of 
\ord{\gamma^2}\/ which must be compensated by some higher order terms. To 
leading order one has
\beq{biast01}
   \frac{1}{S_{01}} 
 = \frac{1}{T_{01}} + \frac{T_{22}}{T^2_{01}} - \frac{T_{23}}{T^3_{01}} 
 + \ord{\gamma^4} .
\eeq
For all other terms the substitution $T_{lm}$\/ for $S_{lm}$\/ is consistent, and 
the variance $\hat{C}(\bar{\mu}_c)$\/ expressed through known quantities becomes
\beq{varmucdat}
    \hat{C}(\bar{\mu}_c) 
 = \frac{1}{T_{01}}  
 + \frac{2 T_{22}}{T^2_{01}}
 - \frac{3 T_{23}}{T^3_{01}}
 + \frac{T^2_{12}}{T^4_{01}}
 + \ord{\gamma^4} .
\eeq
Note that in case of symmetric errors only the first term contributes, recovering 
the conventional estimate for the variance of a weighted average. 

Finally, the expectation value of the $\chi^2$-test-variable shall be examined 
for a weighted average based on measurements with asymmetric errors. Defined 
in the usual way, $\chi^2$\/ is given by
\beq{chi2asym}
 \chi^2 = \sum_k \frac{(\hat{\mu}_k - \bar{\mu})^2}{\hat{C}_k}
        = Z_2 - \frac{Z^2_1}{Z_0}
\eeq
where $\hat{\mu}_k$\/ and $\hat{C}_k$\/ are the unbiased estimates for mean 
value and variance of the individual measurements and $\bar{\mu}$\/ determined
according to \eq{wavbar}. Up to second order the determination of the expectation 
value is easily performed, giving the result
\beq{expchi2}
   \av{\chi^2} 
 = (n-1) + \frac{3}{2} S_{21} 
         - \frac{S^2_{11}+4S_{22}}{S_{01}} 
         + \frac{4 S_{11}S_{12} + 7 S_{23}}{2 S^2_{01}}
         - \frac{2 S^2_{12}}{S^3_{01}} 
   + \ord{\gamma^4},
\eeq 
where $n$\/ is the number of measurements contributing to the weighted average.
For practical applications this result implies, that also in the presence of 
asymmetric errors the usual $\chi^2$-variable can be used as a goodness-of-fit
criterion. For a consistent set of measurements the value should be not too
deviant from the number of degrees of freedom. One should also note the somewhat 
surprising result that a meaningful $\chi^2$-variable is based on the 
non-bias-corrected average $\bar{\mu}$.

Before moving on, it is worth while to discuss the explicit expressions when 
averaging $n$\/ data points which all are drawn from the same parent distribution, 
i.e. $C_k=C$\/ and $\gamma_k=\gamma$\/ for all indices $k$, as then the 
inherent structure of the results becomes most evident. One finds the following
expressions for the expectation values discussed above:
\begin{eqnarray}
    \left\langle \bar{\mu} \right\rangle
 & = & \mu - \gamma \left(1-\frac{1}{n} \right) + \ord{\gamma^3}   \\
    \left\langle C(\bar{\mu}) \right\rangle
 & = & \frac{C}{n} + \frac{\gamma^2}{n} \left(1-\frac{1}{n} \right) + \ord{\gamma^4}   \\
    \av{\chi^2} 
 & = & (n-1) + \frac{\gamma^2}{2C} \left(n - 4 + \frac{3}{n}\right) + \ord{\gamma^4}
\end{eqnarray}
A simple cross check is obtained by setting $n=1$\/ in which case the values 
$\av{\bar{\mu}}=\mu$, $\av{C(\bar{\mu})}=C$\/ and $\av{\chi^2}=0$\/ are recovered 
correctly. One also sees that for $\gamma=0$\/ the usual expressions known for 
weighted averages are obtained. The corrections which are needed for non-linear
relations between measurements and model parameters are an expansion in the 
non-linearity parameter $\gamma$\/ and $1/n$.

\subsection{A Numerical Example}
Since some of the above results may seem counterintuitive they shall be illustrated
by a numerical example. The model assumes that the relation between a true parameter 
$\mu_t$\/ and the expectation value $\av{x}$\/ of a measurement $x$\/ is given by 
$\av{x}=\sqrt{\mu_t}$. Assuming that an experiment provides an unbiased measurement 
$x$\/ with a Gaussian PDF around the mean, the likelihood function is given by 
\beq{likelyx}
    p(x|\mu) \sim \exp\left(-\frac{(x-\sqrt{\mu})^2}{2\sigma^2}\right) .
\eeq
To study the averaging procedure, measurements $x$\/ were generated in a Monte 
Carlo simulation and treated in the usual way to obtain a Maximum Likelihood 
estimate $\mu_{ml}$\/ with asymmetric errors covering a 68\%~confidence level 
interval. The properties of single estimates as well as weighted averages will 
be discussed below.

For the simulation it was assumed that the true value is $\mu_t=25$\/ and that 
the variance of the measurements is $\sigma^2=0.5$. The characteristics of the 
problem then are given by $\av{x}=5$, $\alpha=10$, $\beta=1$, $\gamma=2$ and $C=50.5$.
A single value $x$\/ yields a maximum likelihood estimate $\hat{\mu}_{ml}=x^2$\/ 
with asymmetric errors $\sigma_{\pm}= 2x\sigma\pm\sigma^2$.  

\begin{figure}[htb]
\begin{center}
  \fplace{0.45}{0.cm}{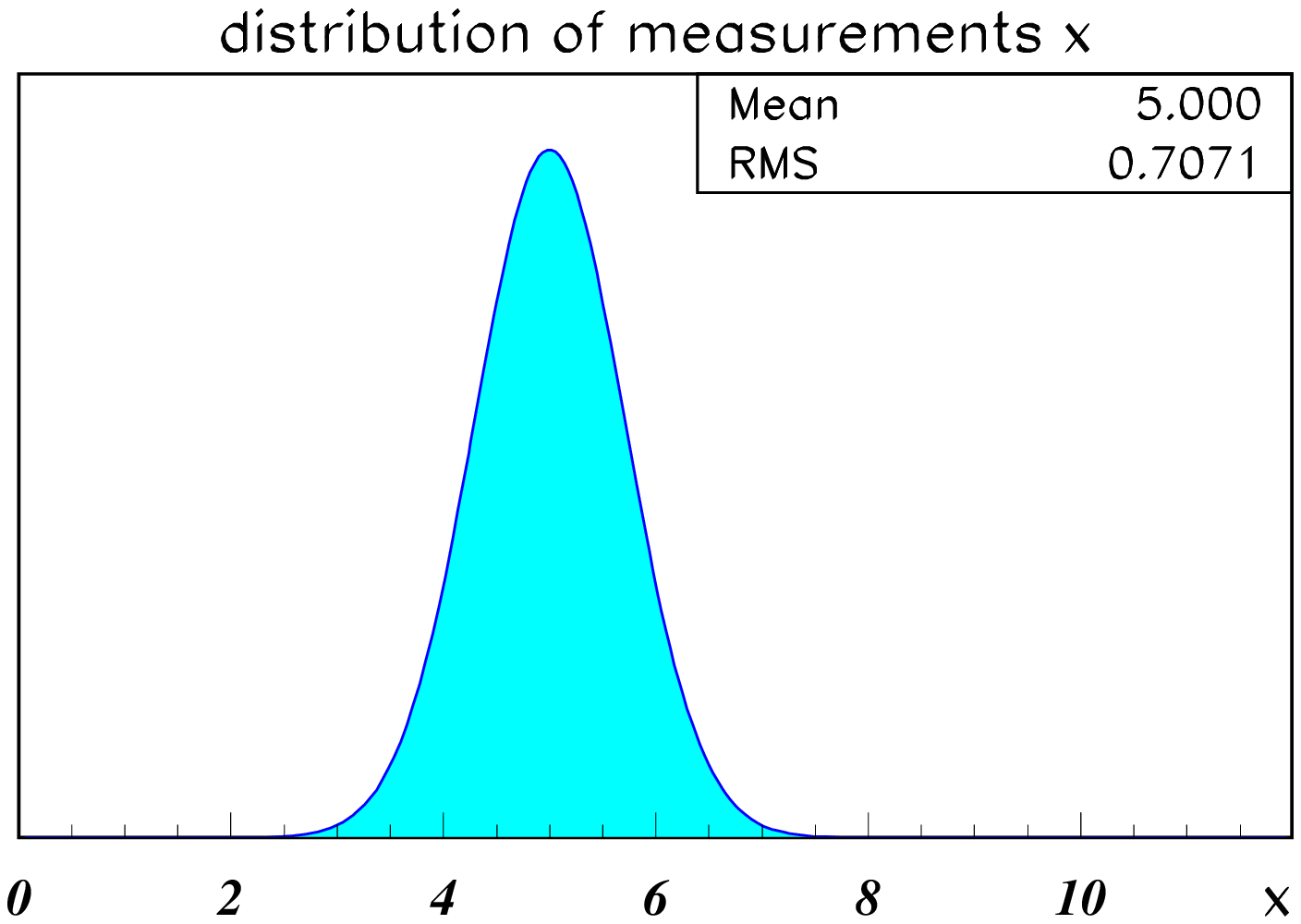} \mm{10}
  \fplace{0.45}{0.cm}{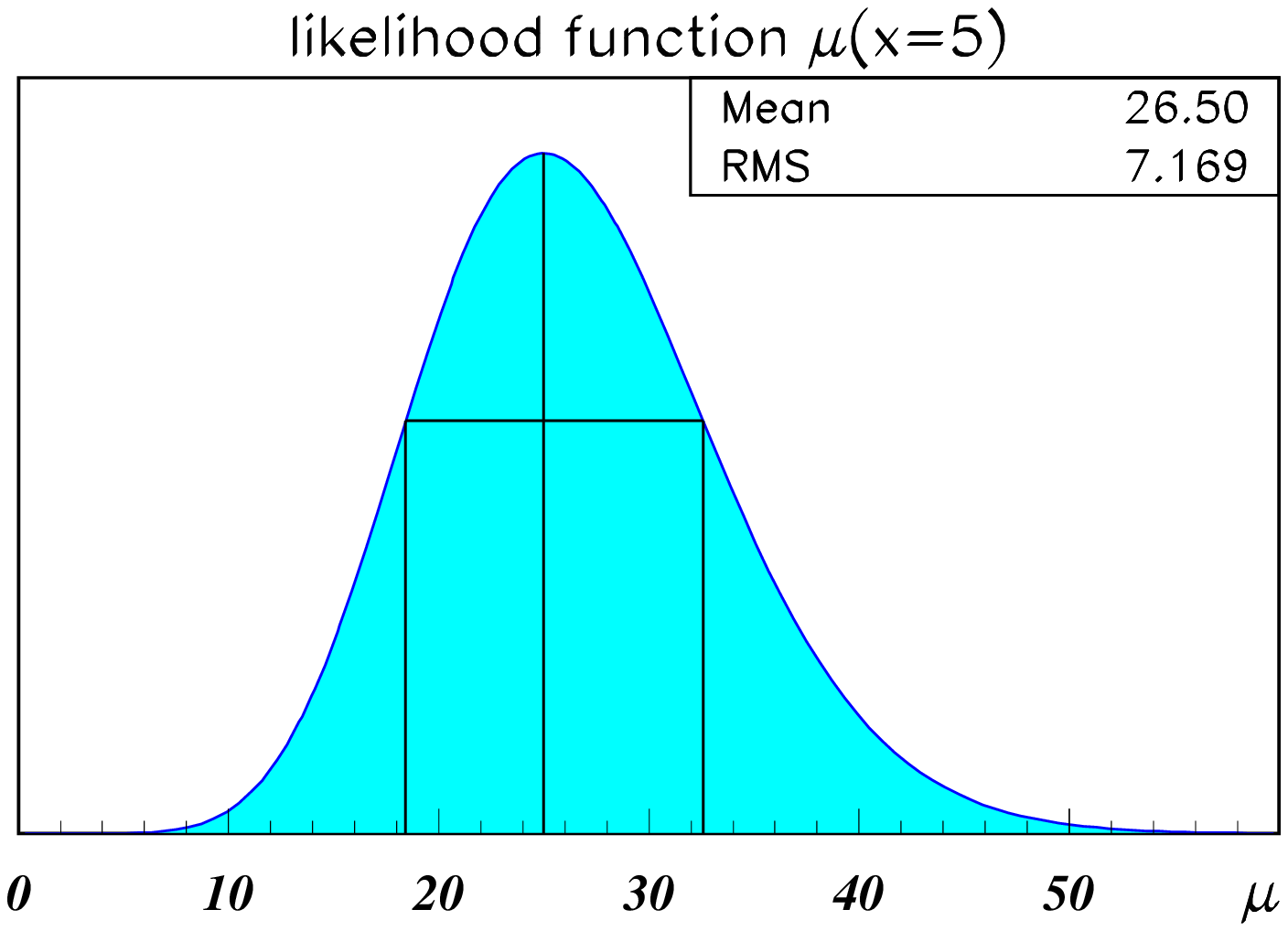} \\[5mm]
  \fplace{0.45}{0.cm}{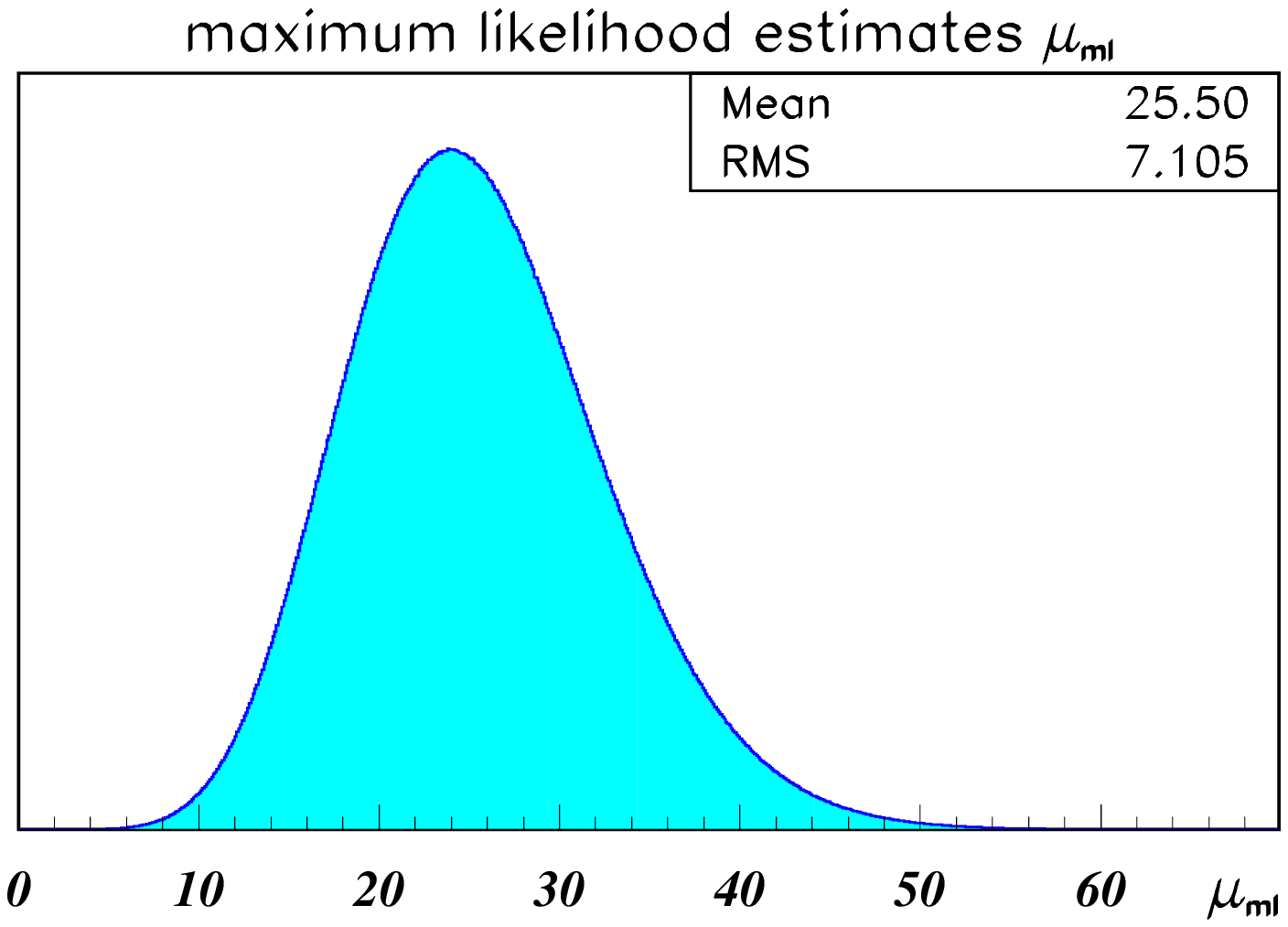} \mm{10}
  \fplace{0.45}{0.cm}{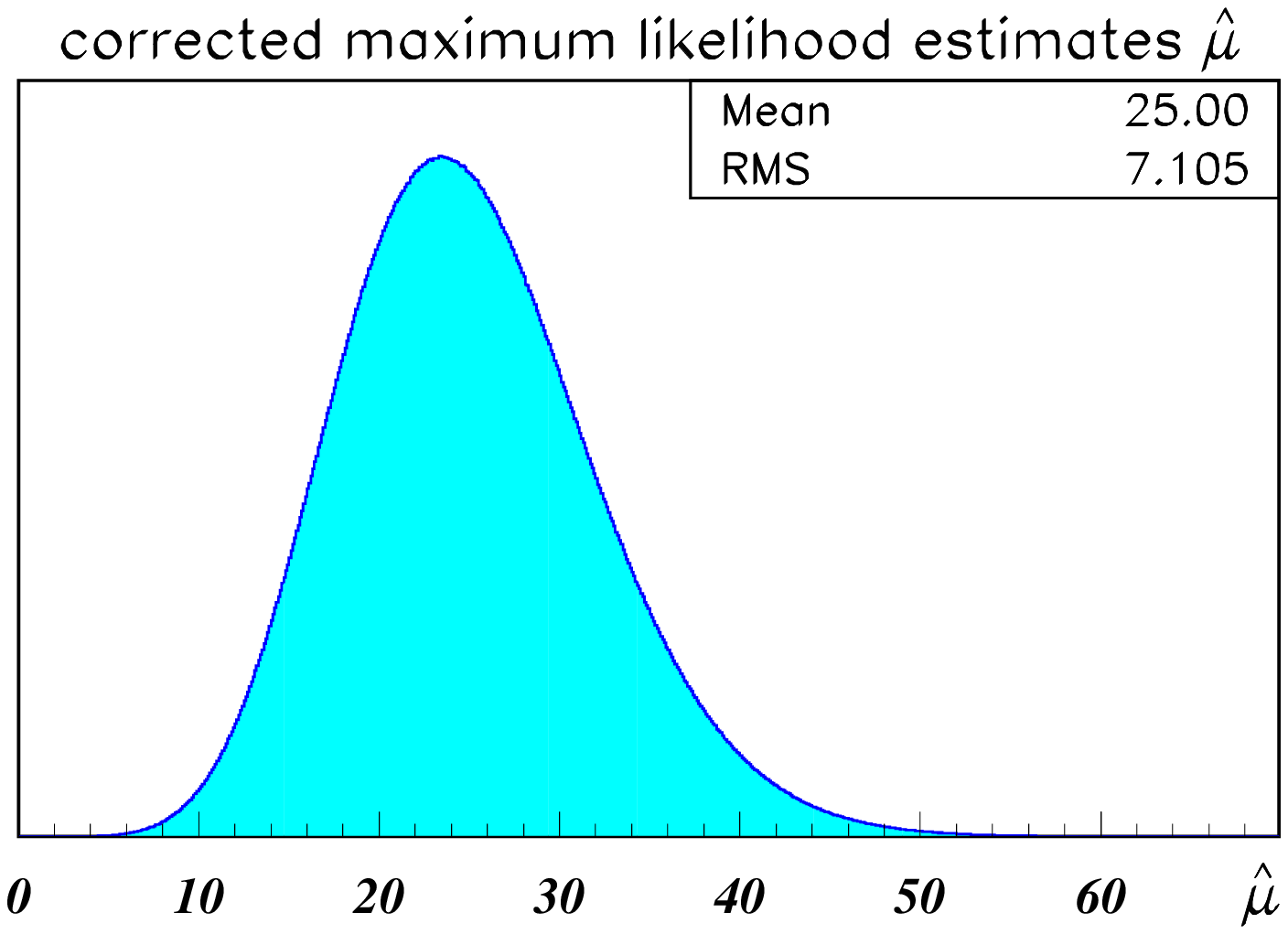}
  \cplace{0.90}{0.cm}{fig:single}
         {Distribution of measurements (upper left), and likelihood function 
          for the parameter $\mu$\/ from a measurement $x=5$\/ (upper right).
          The lower plots show the distribution of the uncorrected (left)
          and the bias corrected (right) Maximum Likelihood estimates, 
          $\mu_{ml}$\/ and $\hat{\mu}$\/ (right), respectively.}
\end{center}
\end{figure}

The distribution of the measurements $x$\/ is shown together with an example for
the slightly asymmetric likelihood function, taking $x=5$, in the top row of 
\fig{single}. For this example the maximum likelihood parameter estimate 
becomes $\mu_{ml}=25\pm^{7.57}_{6.57}$. The two lower plots of \fig{single} 
display the distributions of the maximum likelihood estimates that result from the
Gaussian distribution in $x$. The left hand plot shows the distribution of the 
plain estimate $\mu_{ml}$, the right hand plot has the bias corrected 
distribution of $\hat{\mu}$. The expectation values are $\av{\mu_{ml}}=25.5$\/
and $\av{\hat{\mu}}=25$. The plain Maximum Likelihood estimate evidently is biased, 
consistent with the expected bias $(\sigma_+-\sigma_-)/2=\sigma^2=0.5$. The example 
nicely illustrates that an asymmetric error $\sigma_+ > \sigma_-$\/ really means that 
the maximum likelihood estimate tends to be larger than the true parameter value. 
To correct for the bias one has to reduce the value by the average asymmetry of the 
errors. 

As an aside, it may be interesting to note that using the mean value $\av{\mu(x)}$\/
of the likelihood function $\mu(x)$\/ rather than the maximum as an estimate for 
the unknown parameter actually makes things worse. If spill-over to negative values 
can be ignored, then one finds $\av{\mu(x)}=x^2+3\sigma^2$, i.e. a bias three 
times larger than the bias of the Maximum Likelihood estimate $\mu_{ml}$.
Also this is illustrated in \fig{single} by the likelihood function $\mu(x=5)$.

The Monte-Carlo simulation shows that the variance of the bias corrected maximum 
likelihood estimate $C(\hat{\mu})$\/ is consistent with the expectation $C=50.5$. 
It is interesting to compare this value to a heuristic estimate for the variance, 
based on the average of the positive and the negative errors. The expectation 
value of this heuristic variance is slightly larger than $C$. One finds 
$\av{(\sigma_++\sigma_-)^2/4} = \av{4\sigma^2 x^2} = 4\sigma^2 (\av{x}^2+\sigma^2)=51$.
The term $(\sigma_+-\sigma_-)^2/2$\/ in \eq{varhat} corrects this bias.

\begin{figure}[htb]
\begin{center}
  \fplace{0.45}{0.cm}{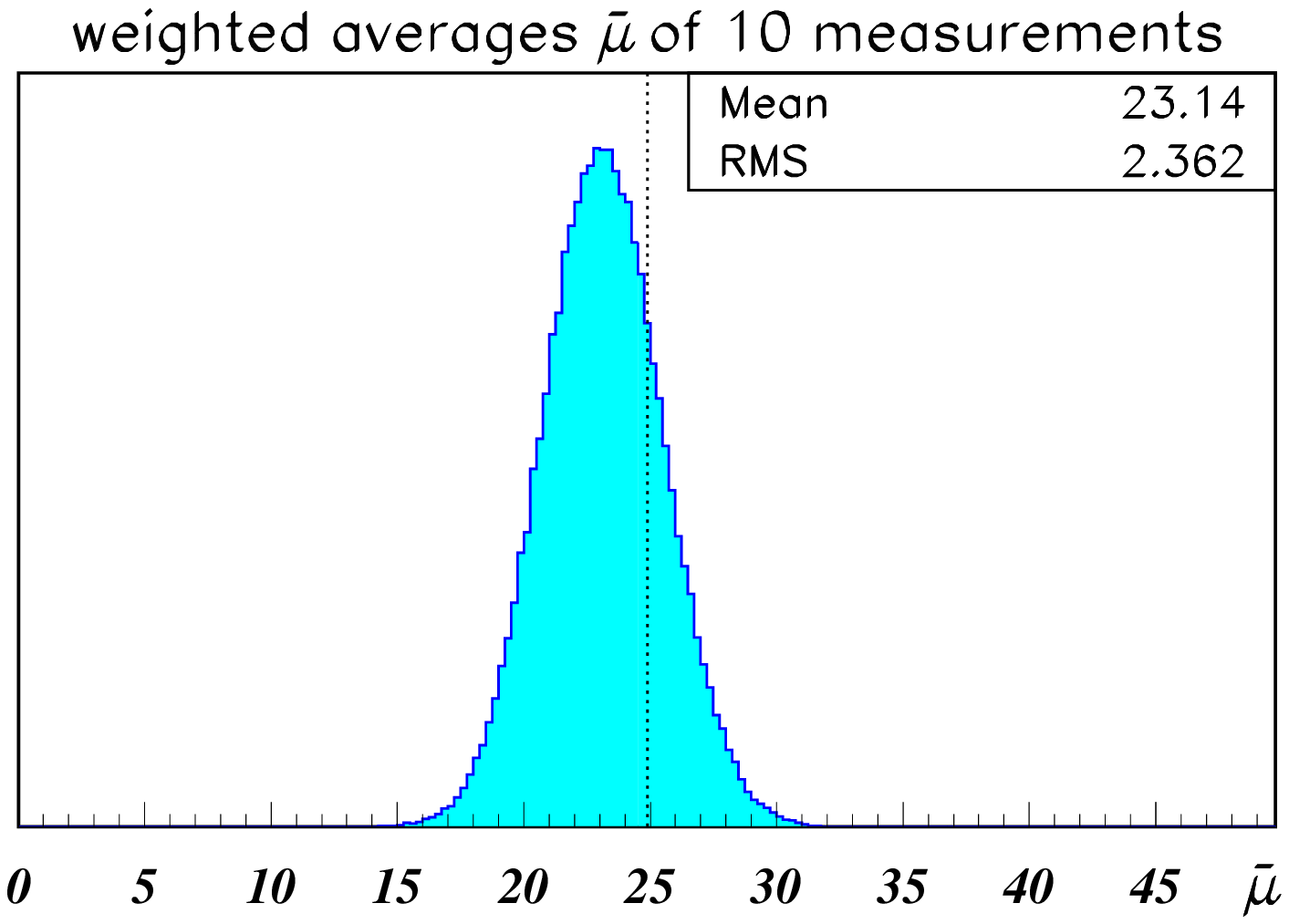} \mm{10}
  \fplace{0.45}{0.cm}{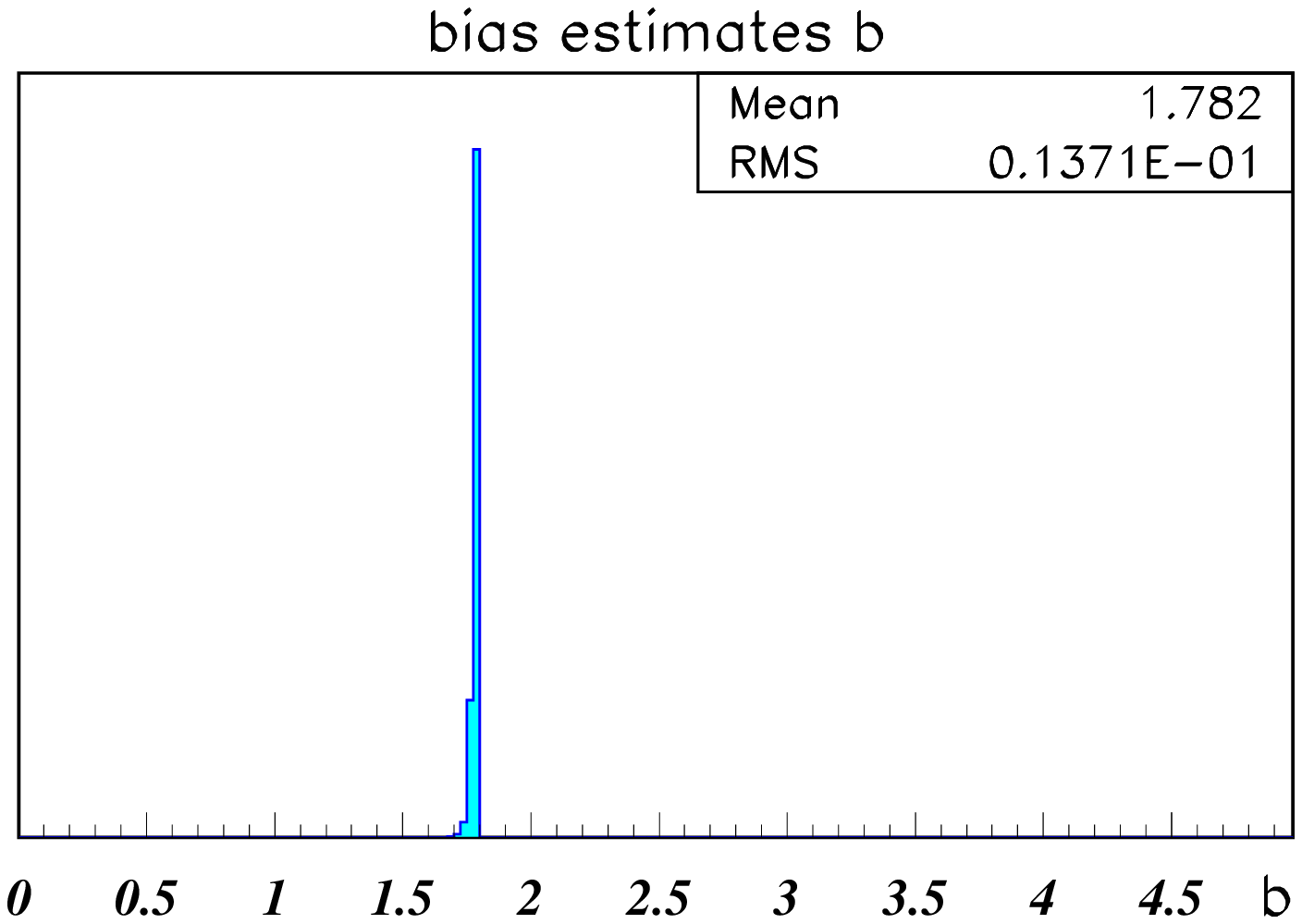} \\[5mm]
  \fplace{0.45}{0.cm}{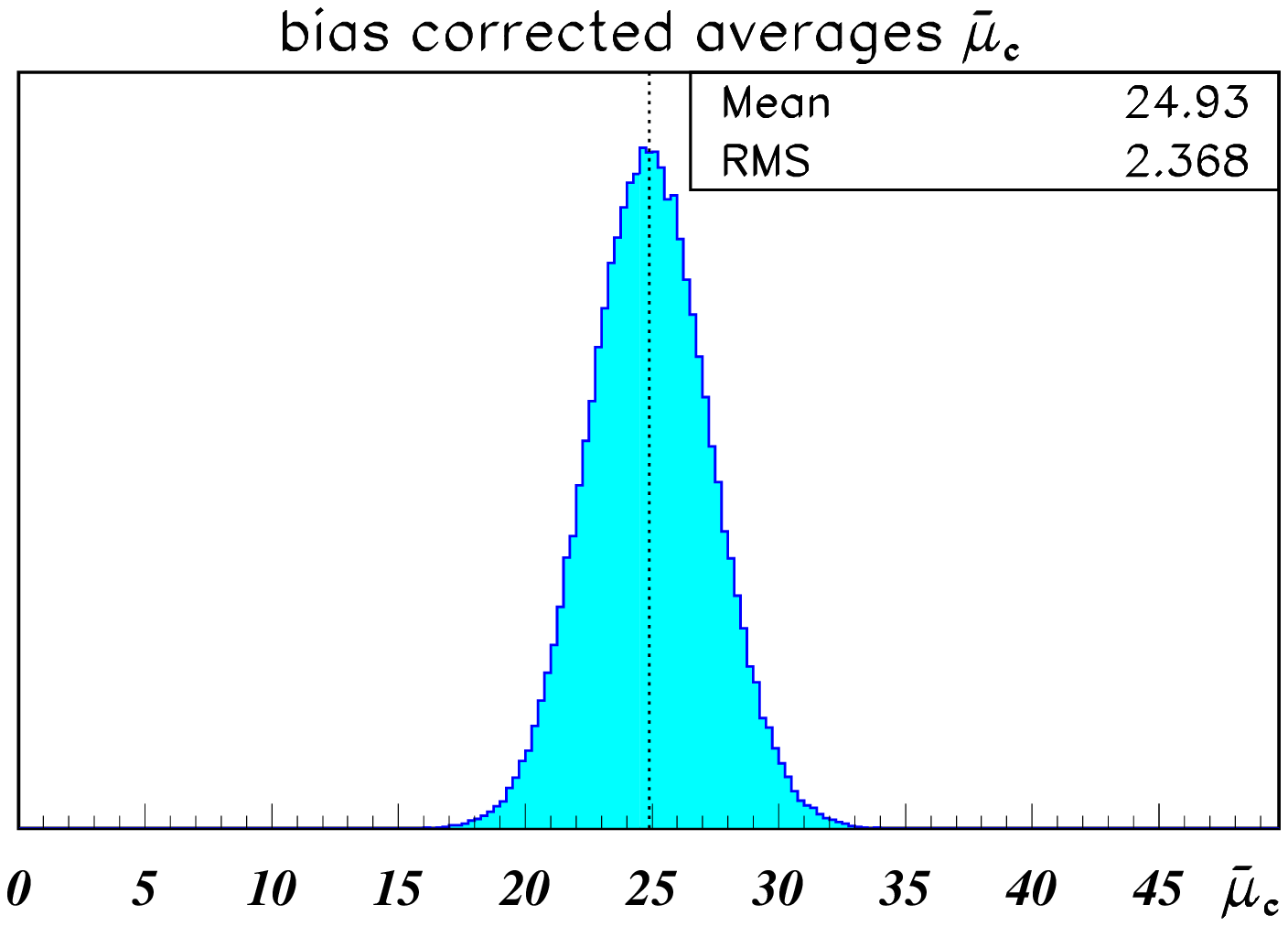} \mm{10}
  \fplace{0.45}{0.cm}{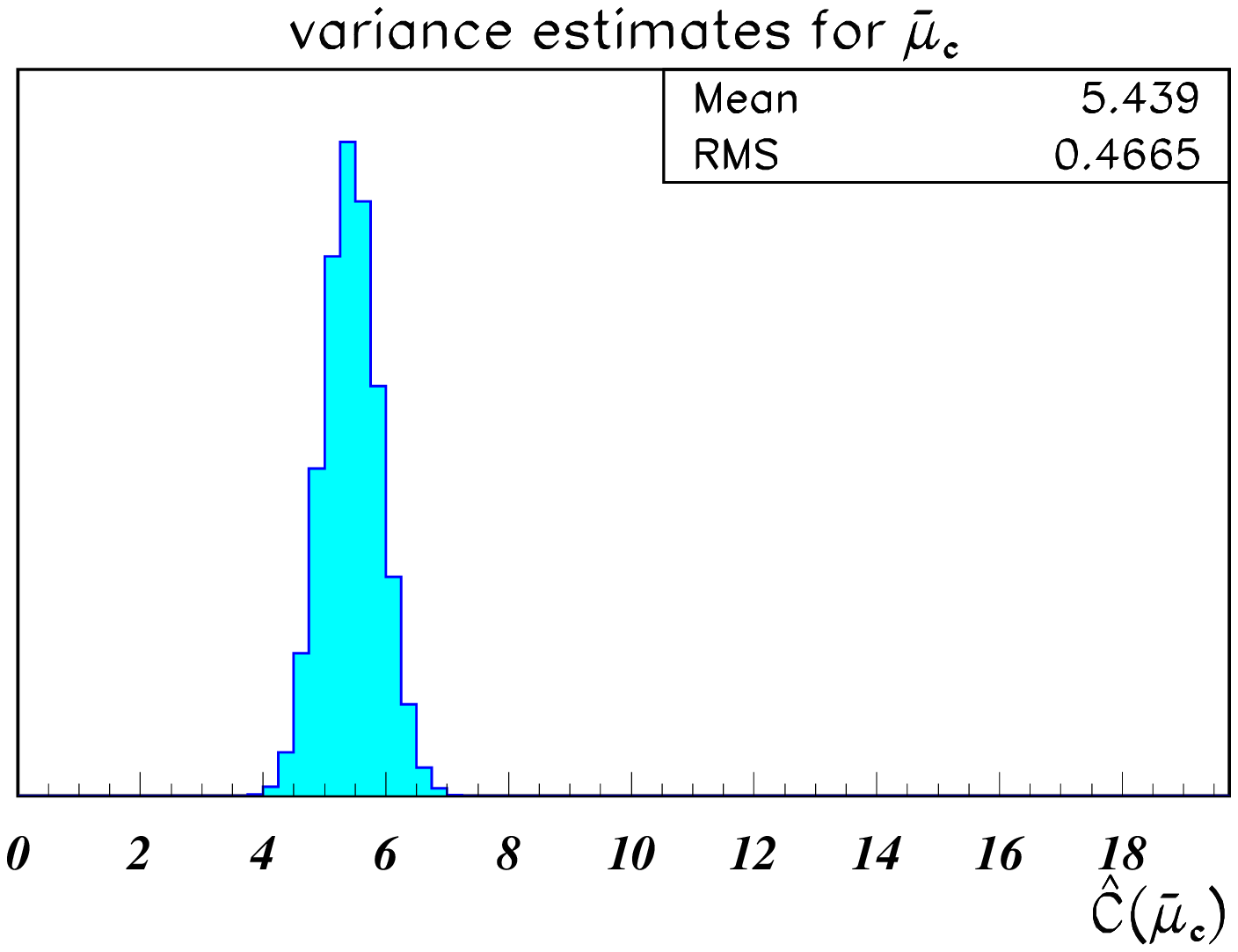}
  \cplace{0.90}{2.mm}{fig:multi}
         {Weighted average of 10 bias-corrected maximum likelihood estimates
          $\hat{\mu}$\/ with unbiased variance estimates $\hat{C}(\hat{\mu})$. 
          The upper left plot shows that the uncorrected average $\bar{\mu}$\/
          is significantly biased. The bias estimate is displayed in the upper 
          right. The distribution of the bias corrected averages $\bar{\mu}_c$\/
          is shown in the lower left, and the distribution of the variance 
          estimates for the corrected averages on the lower right.}
\end{center}
\end{figure}

Having established unbiased estimates for $\mu$\/ and the variance of that 
estimate, the next step is to study the weighted average of results from 
independent measurements. As discussed before, using the inverse of the variance 
estimate as the weight, biases the average because of the correlation between
parameter and variance estimate. The result of using \eq{wavbar} for an average
of 10 values is displayed in the upper left plot of \fig{multi}. The central 
value of the average is about 0.8~standard deviations lower than the true 
parameter value. The estimated bias is shown in the upper right. According to 
\eq{sumbias}, i.e. neglecting terms of \ord{\gamma^4}, one would expect a shift 
of 1.8 units, which is close to what is actually observed. The bias corrected 
average is shown in the lower left of \fig{multi}. The lower right finally 
shows the distribution of the variance estimate for the bias corrected weighted
average. The distribution is rather narrow with a central value close to the 
actual variance found in the distribution of the averages. That the general 
picture is consistent is also shown in the $\chi^2$-distribution of the weighted
average, \fig{fchi2}. According to \eq{expchi2}, including all terms up to 
\ord{\gamma^2} one would predict an expectation value $\av{\chi^2}\approx 9.25$,
again very close to the observed value.

\begin{figure}[htb]
\begin{center}
  \fplace{0.45}{0.cm}{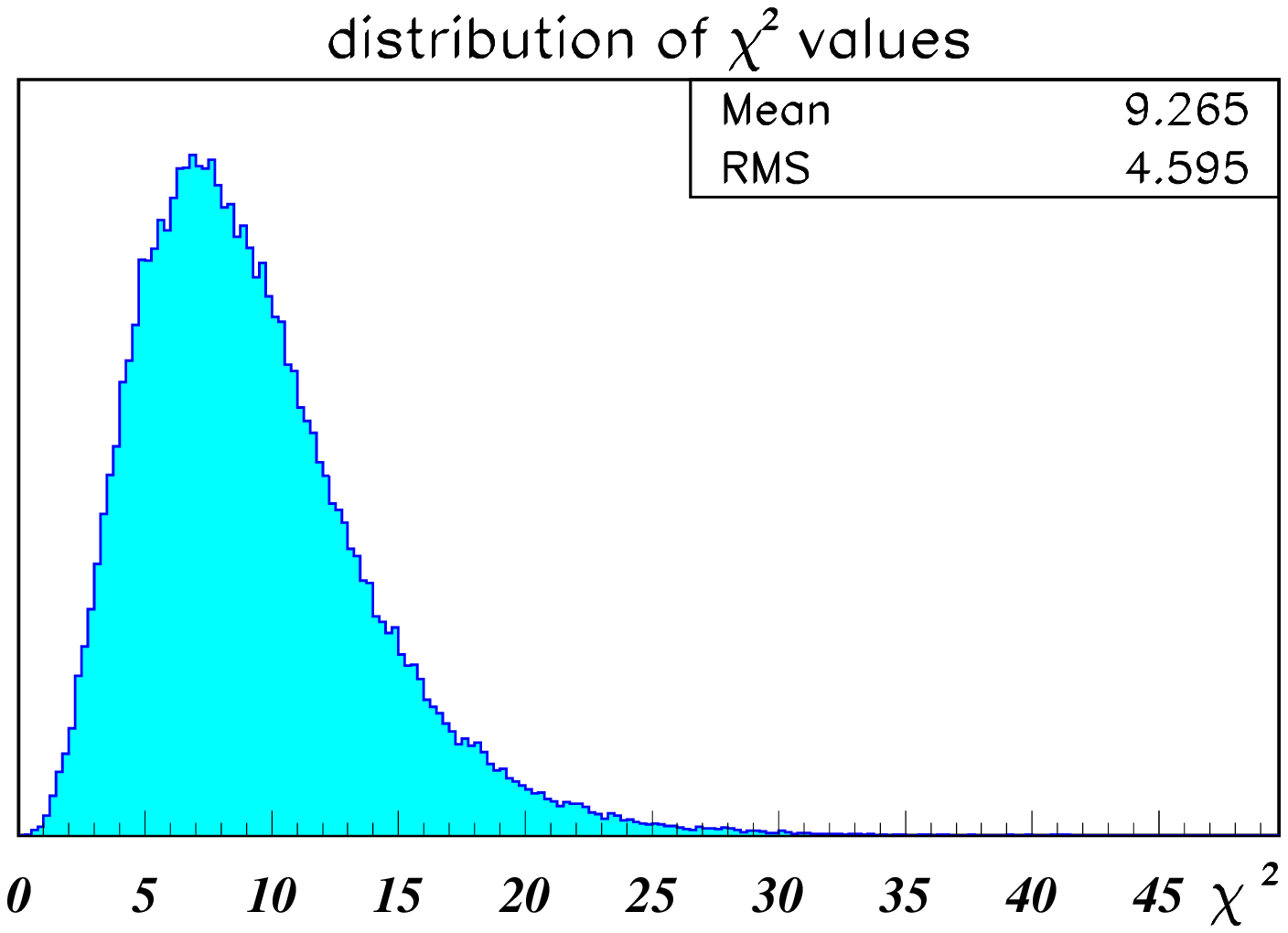} \hfill
  \cplace{0.45}{2.cm}{fig:fchi2}
         {Distribution of $\chi^2$-values for the weighted average of 
          10 bias-corrected maximum likelihood estimates.}
\end{center}
\end{figure}

To illustrate further the properties of different schemes to form a weighted 
average from measurements with asymmetric errors, \fig{fevol} shows how the 
various methods converge when increasing the number $n$\/ of measurements. Here
$n$\/ was varied from 1~to~50. The top plot shows the result of a simple 
heuristic method, where the maximum likelihood estimates $\mu_{ml}$\/ are 
averaged, using as weights the inverse of $(\sigma_++\sigma_-)^2/4$, the middle 
diagram was obtained by using unbiased, though correlated, estimates for 
mean value and variance, and the lower plot shows the result of the proper 
bias correction scheme. The first two schemes yield very similar results, the 
main difference being a shift corresponding to the bias correction for an 
individual measurement. The heuristic method has a large bias for $n=1$, which 
is corrected in the second method. At large $n$\/ both significantly underestimate
the true parameter value. One also has to mention that both provide error 
estimates which underestimate the actual uncertainty of an individual weighted 
average by roughly 10\%. The proper bias correction scheme has both a stable 
expectation value for the average and a reliable error estimate.

\begin{figure}[htb]
\begin{center}
  \fplace{0.50}{0.cm}{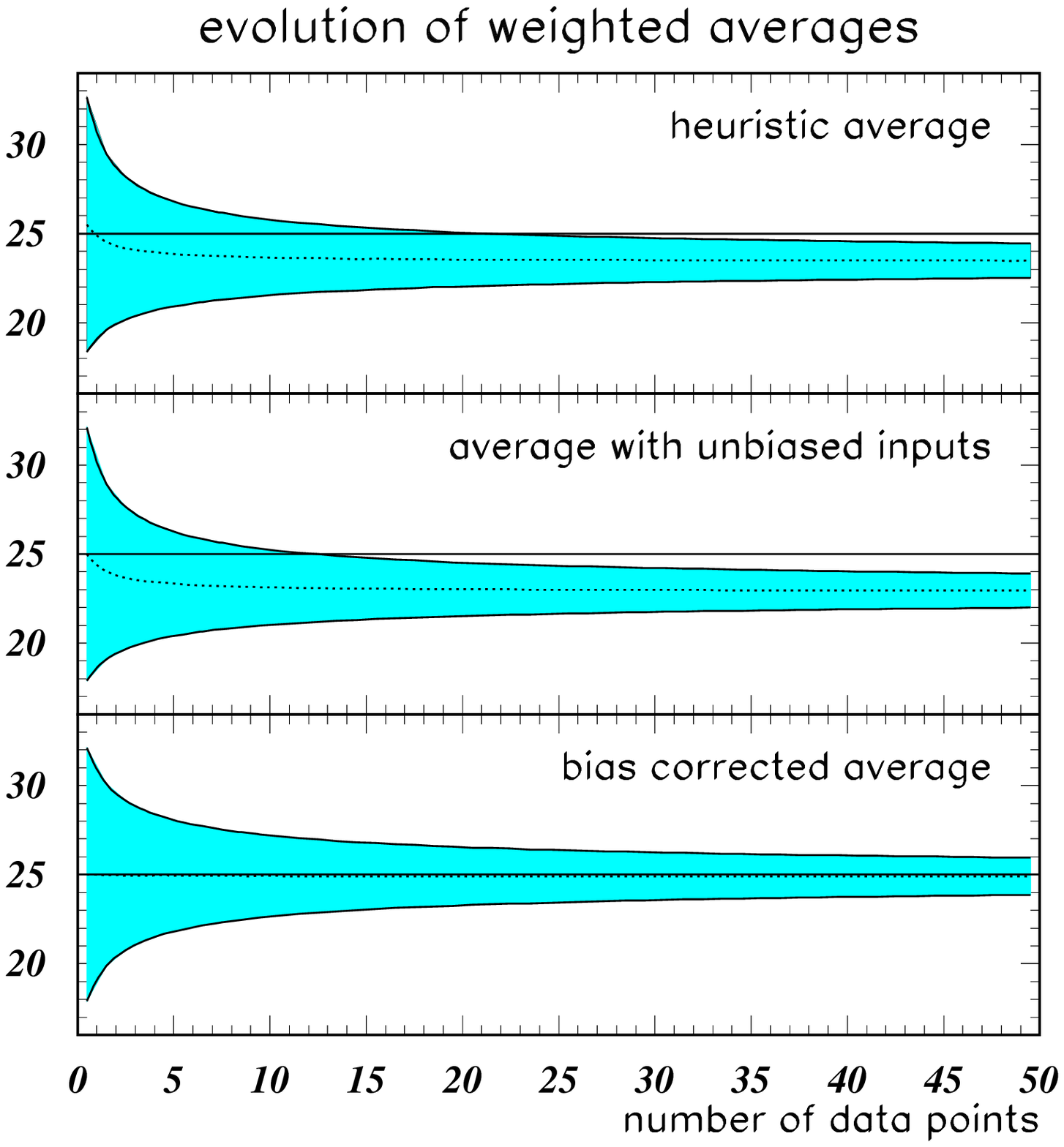} \mm{10}
  \cplace{0.40}{1.cm}{fig:fevol}
         {Evolution of mean value and error estimate as function of 
          the number of measurements in the average. The top plot shows
          the behavior of a simple heuristic treatment of asymmetric 
          errors, the middle one the result of using bias corrected 
          estimates $\hat{\mu}$\/ and $\hat{C}(\hat{\mu})$\/ for mean
          value and variance, and the lower plot the results after 
          proper bias correction for the average and its variance.}
\end{center}
\end{figure}

\section{Systematic Uncertainties}
The previous considerations apply to statistical uncertainties. Many measurements 
are also subject to systematic or theoretical uncertainties, which affect the 
model that is fitted to the experimental data. Formally the effect of these 
uncertainties can be described by an additional parameter $a$\/ entering the 
model, where the knowledge about $a$\/ is described by a PDF $h(a)$. The function 
$h(a)$\/ can be the result of a measurement or a purely Bayesian parameterization 
for the subjective degree of belief in a certain value $a$. The formal treatment
will be unaffected by this. With the parameter $a$\/ \eq{pxmu} becomes
\beq{pxmua}
   p(x|\mu,a) = \frac{1}{\sqrt{2\pi}\sigma}
                \exp\left( -\frac{(x-s(\mu,a))^2}{2\sigma^2} \right). 
\eeq         

The simplest and probably still very common approach to deal with the situation
is to vary $a$\/ according to $h(a)$\/ around a nominal value $a_0$\/ over a 
range covering a 68\% confidence level interval, and quote the resulting variation 
in $\mu$\/ as a systematic error. The discussion presented here will deal with 
only this approach. As before, from the quoted errors the mean value and variance
of the parameter estimate shall be inferred, since that is the information which 
is needed to form a weighted average.

For a fixed parameter $a$\/ the maximum likelihood estimate $\mu_{ml}$\/ for 
$\mu$\/ according to \eq{pxmua} is given by $\mu_{ml}=s^{-1}(x,a)$. The resulting 
PDF $\rho(\mu_{ml})$\/ for $\mu_{ml}$\/ depends on $s^{-1}(x,a)$\/ and the PDF 
$h(a)$\/ of $a$.
\beq{rhomuhat}
  \rho(\mu_{ml}) = \int da\; h(a) \; \delta\left( \mu_{ml} - s^{-1}(x,a) \right).
\eeq
From \eq{rhomuhat} one finds for the moments of $\mu_{ml}$
\beq{mommuhat}
 \av{\mu_{ml}^k} = \int da\; h(a)\; \left( s^{-1}(x,a) \right)^k .
\eeq
To continue, it will be assumed that $h(a)$\/ is a Gaussian with mean value $a_0$\/ 
and standard deviation $\tau$, and that $s^{-1}(x,a)$\/ has an expansion around 
$a_0$\/ 
\beq{expandsxa}
   s^{-1}(x,a) = \mu_0(x) + \alpha(x) (a-a_0) + \beta(x)(a-a_0)^2
\eeq
where the higher order terms are negligible. To simplify the notation the 
$x$-dependence of the coefficient will be dropped. With 
\beq{hagauss}
  h(a) = \frac{1}{\sqrt{2\pi} \tau} \exp\left( -\frac{(a-a_0)^2}{2\tau^2} \right)
\eeq
one finds for the first and second moment of $\mu_{ml}$
\beq{moms}
   \av{\mu_{ml}} = \mu_0 + \beta\tau^2 
   \mm{5} \mbox{and} \mm{5} 
   \av{\mu_{ml}^2} =  \mu_0^2 + (\alpha^2+2\mu_0\beta)\tau^2 + 3\beta^2\tau^4 . 
\eeq
From \eq{moms} one finds the variance of $\mu_{ml}$
\beq{varmuhatsys}
    C(\mu_{ml}) 
 = \av{\mu_{ml}^2} - \av{\mu_{ml}}^2
 = \alpha^2\tau^2 + 2\beta^2\tau^4 .
\eeq
For $a=a_0$\/ one has the nominal parameter estimate $\mu_0$. The in general asymmetric
errors are given by 
\beq{asymerr}
   \tau_+ = \alpha\tau + \beta\tau^2
   \mm{5} \mbox{and} \mm{5}  
   \tau_- = \alpha\tau - \beta\tau^2 . 
\eeq
Combining the information from \eq{moms} and \eq{asymerr} the unbiased parameter 
estimate and its variance can be written as
\beq{sysfinal}
  \av{\mu_{ml}} = \mu_0 + \frac{1}{2}(\tau_+ - \tau_-) 
  \mm{5} \mbox{and} \mm{5}  
  C(\mu_{ml}) = \frac{1}{4} (\tau_+ + \tau_-)^2 + \frac{1}{2} (\tau_+ - \tau_-)^2.
\eeq
These expressions \eq{sysfinal} show how to deal with theoretical or systematic 
uncertainties when averaging independent measurements. In contrast to the statistical 
errors discussed before, now asymmetric errors translate into a shift of the parameter
as one would naively expect. Also the variance is strictly positive. Although the 
derivation is formally very similar to the one of the statistical errors, here the 
PDF of the function of the parameter $a$\/ is known a priori, which is not the case
for the measurement $x$. Note that the expressions \eq{sysfinal} have been derived 
for the assumption that systematic uncertainties are Gaussian errors.

\section{Combination of Statistical and Systematic errors}
Except in the context of Bayesian statistics the combination of statistical and 
systematic errors is not a well defined concept. If one, however, adopts the 
attitude to model the influence of the two kinds of uncertainties by means of a
Monte Carlo method, where the measurement $x$\/ as well as the theory-parameter
$a$\/ is varied according to their respective probability density functions, 
then the total bias and variance for the parameter estimate $\hat{\mu}$\/ is 
given by the sum of the individual biases and the individual variances, i.e.
the errors can be combined in quadrature: 
\beq{statsyst}
  C_k= C_k^{\mbox{stat}} + C_k^{\mbox{syst}}.
\eeq
In the Bayesian sense this is the correct way of combining different sources
of uncertainty. From a non-Bayesian point of view the ansatz \eq{statsyst} 
at least has the qualitatively desired properties, that measurements with 
large uncertainty in either of the two components will get a low weight in 
any average.

When using the combined variances in a weighted average, one again needs to 
perform a bias correction based on the asymmetry parameters $\gamma$. As 
defined in \eq{varbias}, only the difference of the statistical errors contributes
to $\gamma$, since only asymmetries in the statistical uncertainties lead  
to a correlation between variance and mean value. No such correlation exists for 
systematic errors. Otherwise the definitions~\eq{deftlm} apply as before.

\section{Summary and Conclusions}
The weighted average with weights based only on the variance of the individual 
measurements is a robust estimator for a common mean. As a consequence of the 
central limit theorem, its PDF can be assumed to be a Gaussian even if the PDFs 
of the inputs are not known in detail. Although it would be preferable from a 
theoretical point of view to take correlations into account when forming the 
average, this can only be recommended if those are well understood. A simple 
effective way is proposed to include the impact of correlations between the 
measurements for the error estimate, using the $\chi^2$\/ of the average as an 
indicator variable. 

In a second step the problem of asymmetric 68\% confidence level error intervals 
was addressed. The argument was based on the assumption of a measurement with a 
Gaussian PDF and the use of the Maximum Likelihood Method to infer the value of a 
physical parameter and its error. Explicit expressions were derived to convert the 
given information into an unbiased estimate for the value of the parameter and its 
variance. In addition, it has been shown that when dealing with asymmetric errors,
correction terms for the weighted average and its variance have to be taken into 
account. In case of symmetric errors the usual expressions for the weighted average 
are recovered. The scaling prescription to account for correlations between 
measurements can be applied in the same way for measurements with symmetric and
asymmetric errors.

Finally, systematic errors were studied, assuming that they enter as model 
parameters with a known PDF that has to be interpreted in the Bayesian sense. 
As such they behave differently from the statistical errors of a measurement. 
Also here explicit expressions are given how to infer mean value and variance and 
how to combine them with statistical uncertainties.

\clearpage
\section*{Acknowledgments}
Sincere thanks go to G\"unther Dissertori and Anke-Susanne M\"uller for
careful reading of the manuscript and for stimulating and constructive 
discussions.


\appendix
\section*{Appendix: Auxiliary Expectation Values}
Here the complete set of expectation values is given which are needed to derive 
the results of section~\ref{sec:wavasym}. In most cases a precision of at least 
\ord{\gamma^2} for the expectation values is obtained already when truncating 
the expansion \eq{expfterms} after the second term.
\beq{trunc}
    \left\langle F \right\rangle 
  = F(0) + \sum_k \frac{\av{d_k^2}}{2!} \frac{\partial^2}{\partial d_k^2}F(0)
    + \ldots
\eeq
Elementary calculus with the derivatives and definitions given in \eq{avwavbar}, 
\eq{deftlm} and \eq{dtlm} then yields the following results:
\begin{eqnarray}
   \av{\frac{1}{T_{01}}} 
 & = & \frac{1}{S_{01}} - \frac{S_{22}}{S^2_{01}} + \frac{S_{23}}{S^3_{01}} 
       + \ord{\gamma^4}                               \\
       \left\langle \frac{T_{11}}{T_{01}} \right\rangle
 & = & \frac{S_{11}}{S_{01}} + \frac{S_{32}}{S_{01}} 
                         - \frac{S_{11}S_{22}+S_{33}}{S^2_{01}}
                         + \frac{S_{11}S_{23}}{S^3_{01}} + \ord{\gamma^5} \\
   \left\langle \frac{T^2_{11}}{T^2_{01}} \right\rangle
 & = &  \frac{S^2_{11}}{S^2_{01}} 
   + \frac{2 S_{11} S_{32} + S_{43} }{S^2_{01}} 
   - \frac{2 S^2_{11} S_{22} + 4 S_{11} S_{33}}{S^3_{01}}
   + \frac{3 S^2_{11} S_{23}}{S^4_{01}} + \ord{\gamma^6}   \\
   \left\langle \frac{T_{12}}{T^2_{01}} \right\rangle
 & = & \frac{S_{12}}{S^2_{01}} + \frac{S_{33}}{S^2_{01}} 
                         - \frac{2S_{12}S_{22}+4S_{34}}{S^3_{01}}
                         + \frac{3S_{12}S_{23}}{S^4_{01}} + \ord{\gamma^5}   \\
   \left\langle \frac{T^2_{12}}{T^4_{01}} \right\rangle
 & = &  \frac{S^2_{12}}{S^4_{01}} 
   + \frac{6 S_{12} S_{33} + 4 S_{45} }{S^4_{01}} 
   - \frac{4 S^2_{12} S_{22} + 16 S_{12} S_{34}}{S^5_{01}}
   + \frac{10 S^2_{12} S_{23}}{S^6_{01}} + \ord{\gamma^6}    \\
   \left\langle \frac{T_{11} T_{12}}{T^3_{01}} \right\rangle
 & = &  \frac{S_{11} S_{12}}{S^3_{01}} 
        + \frac{S_{12} S_{32} + 3 S_{11} S_{33} + 2 S_{44}}{S^3_{01}}  \\
 &   &  - \frac{3 S_{11} S_{12} S_{22} + 3 S_{12} S_{33} + 6 S_{11} S_{34}}{S^4_{01}}
        + \frac{6 S_{11} S_{12} S_{23} }{S^5_{01}} + \ord{\gamma^6}    \nonumber  \\
    \left\langle \frac{Z_1}{Z_0} \right\rangle
 & = &-\frac{S_{11}}{S_{01}} + \frac{S_{12}}{S^2_{01}} + \ord{\gamma^3} \\
    \left\langle \frac{T_{11} Z_1}{T_{01} Z_0} \right\rangle
 & = & -\frac{S^2_{11} + S_{22}}{S^2_{01}} + \frac{2 S_{11} S_{12}}{S^3_{01}}
       + \ord{\gamma^4}  \\
    \left\langle \frac{T_{12} Z_1}{T^2_{01} Z_0} \right\rangle
 & = &-\frac{S_{11} S_{12} + 2 S_{23}}{S^3_{01}} + \frac{3 S^2_{12}}{S^4_{01}}
      + \ord{\gamma^4}
\end{eqnarray}
The quantities where one has to go to fourth order in \eq{expfterms} in order
to collect all terms up to \ord{\gamma^2}\/ are $\av{Z^2_1/Z^2_0}$, $\av{Z^2_1/Z_0}$\/
and $\av{Z_2}$. For $F=Z^2_1/Z^2_0$\/ the partial derivatives up to fourth order are: 
\begin{eqnarray}
       \frac{1}{2!} \frac{\partial^2}{\partial d_k^2} F(0)
 & = & \frac{1}{S^2_{01}}\frac{1}{C^2_k}                    \\
       \frac{1}{3!} \frac{\partial^3}{\partial d_k^3} F(0) 
 & = & - \frac{1}{S^2_{01}}\frac{2\gamma_k}{C^3_k}
       + \frac{1}{S^3_{01}}\frac{2\gamma_k}{C^4_k}          \\
       \frac{1}{4!} \frac{\partial^4}{\partial d_k^4}\ F(0)
 & = & \frac{1}{S^2_{01}}\frac{3\gamma^2_k}{C^4_k}
       - \frac{1}{S^3_{01}}\frac{6\gamma^2_k}{C^5_k}
       + \frac{1}{S^4_{01}}\frac{3\gamma^2_k}{C^6_k}        \\
    \frac{1}{2!2!} 
    \frac{\partial^2}{\partial d_k^2}\frac{\partial^2}{\partial d_l^2} F(0)
 & = &  \frac{1}{S^2_{01}}\frac{2\gamma_k\gamma_l}{C^2_k C^2_l}  
   - \frac{1}{S^3_{01}}\left(  \frac{2\gamma^2_k}{C^3_k C^2_l}   
                             + \frac{4\gamma_k\gamma_l}{C^2_k C^3_l}  
                             + \frac{4\gamma_k\gamma_l}{C^3_k C^2_l} 
                             + \frac{2\gamma^2_l}{C^2_k C^3_l} \right)  \\ 
 &   &  + \frac{1}{S^4_{01}}\left(  \frac{3\gamma^2_k}{C^4_k C^2_l}   
                                  + \frac{12\gamma_k\gamma_l}{C^3_k C^3_l}   
                                  + \frac{3\gamma^2_l}{C^2_k C^4_l} \right)  \nonumber
\end{eqnarray}
With the moments given in \eq{dkmoments} and using the identity
\beq{sumprod}
   2 \sum_{k<l} a_k b_l 
 = \left(\sum_k a_k\right)\left(\sum_l b_l\right) - \left(\sum_k a_k b_k\right)
\eeq
the expectation value $\av{Z^2_1/Z^2_0}$\/ finally becomes
\beq{biasr9}
    \av{\frac{Z^2_1}{Z^2_0}} 
 =   \frac{1}{S_{01}}
   + \frac{S^2_{11} + 3 S_{22}}{S^2_{01}} 
   - \frac{4 S_{11} S_{12} + 6 S_{23}}{S^3_{01}} 
   + \frac{6 S^2_{12}}{S^4_{01}} 
   + \ord{\gamma^4}.
\eeq
Similarly one finds for the other expectation values
\begin{eqnarray}
  \av{\frac{Z^2_1}{Z_0}} 
 & = & 1 + \frac{S^2_{11} +4 S_{22}}{S_{01}}
     - \frac{4 S_{11} S_{12} + 7 S_{23}}{2 S^2_{01}}
     + \frac{2 S^2_{12}}{s^3_{01}} 
     + \ord{\gamma^4}   \\
  \av{Z_2} 
 & = & n + \frac{3}{2} S_{21} + \ord{\gamma^4} 
\end{eqnarray}
where $n$\/ is the number of measurements contributing to the average.

\end{document}